\documentclass{ws-procs9x6}            

\begin{document}

\title{Rotating Hot White Dwarfs}

\author{K. A. Boshkayev,$^{1,2}$ J. A. Rueda$^{1,2}$ and B. A. Zhami$^{1,*}$}

\address{$^1$Department of Physics and Technology, Al-Farabi Kazakh National University,\\
Al-Farabi avenue 71, Almaty, 050040, Kazakhstan\\
$^2$International Center for Relativistic Astrophysics Network,\\
Piazza della Repubblica 10, Pescara, I-65122, Italy\\
$^*$E-mail: zhami.bakytzhan@gmail.com}

\begin{abstract}
We consider the effects of rotation and temperature on the structure of white dwarfs in order to compare them with the estimated data from observations.  
\end{abstract}
\bodymatter

\

\begin{figure}
\begin{center}
\includegraphics[width=0.45\textwidth]{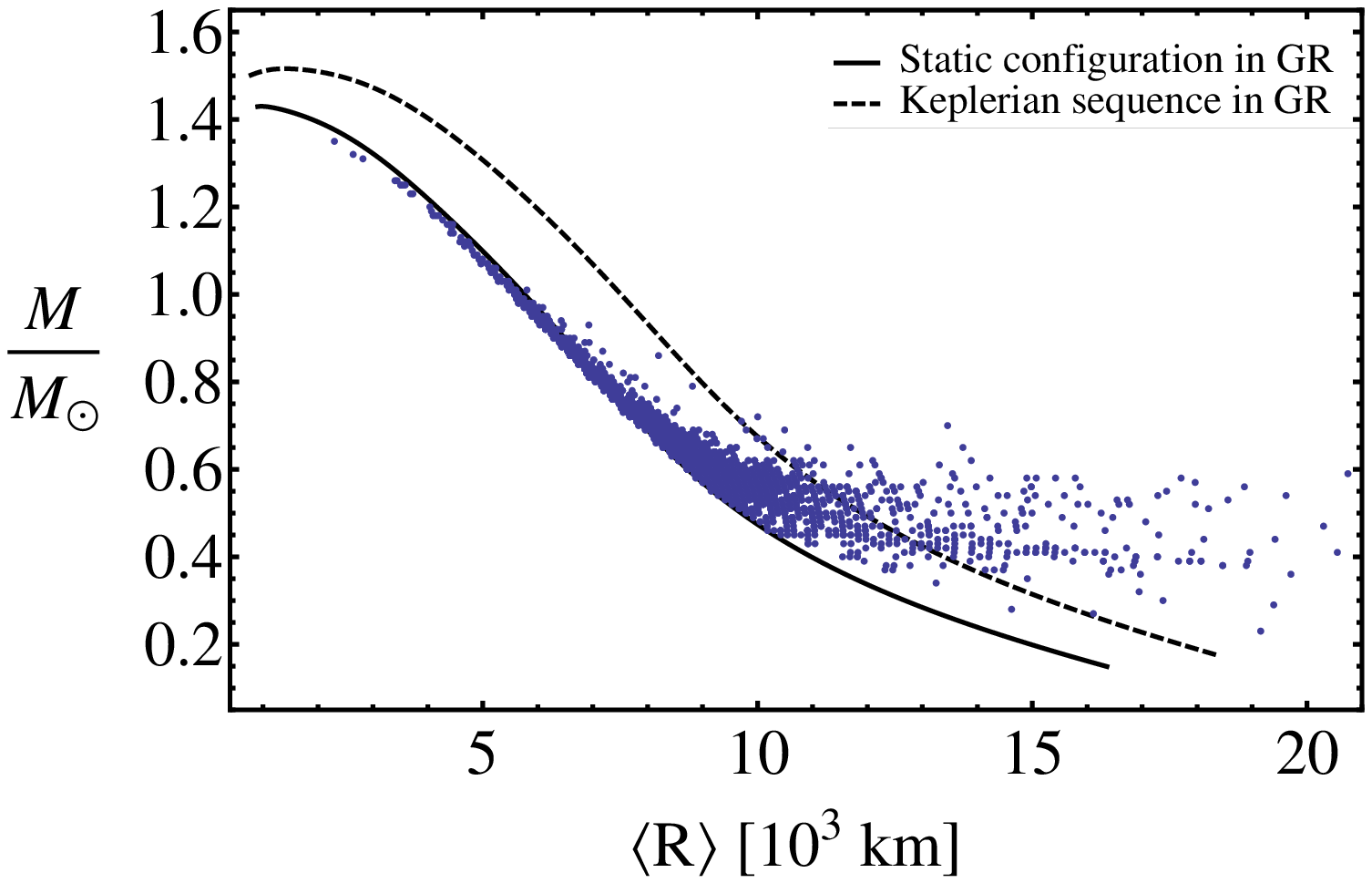}\qquad
\includegraphics[width=0.45\textwidth]{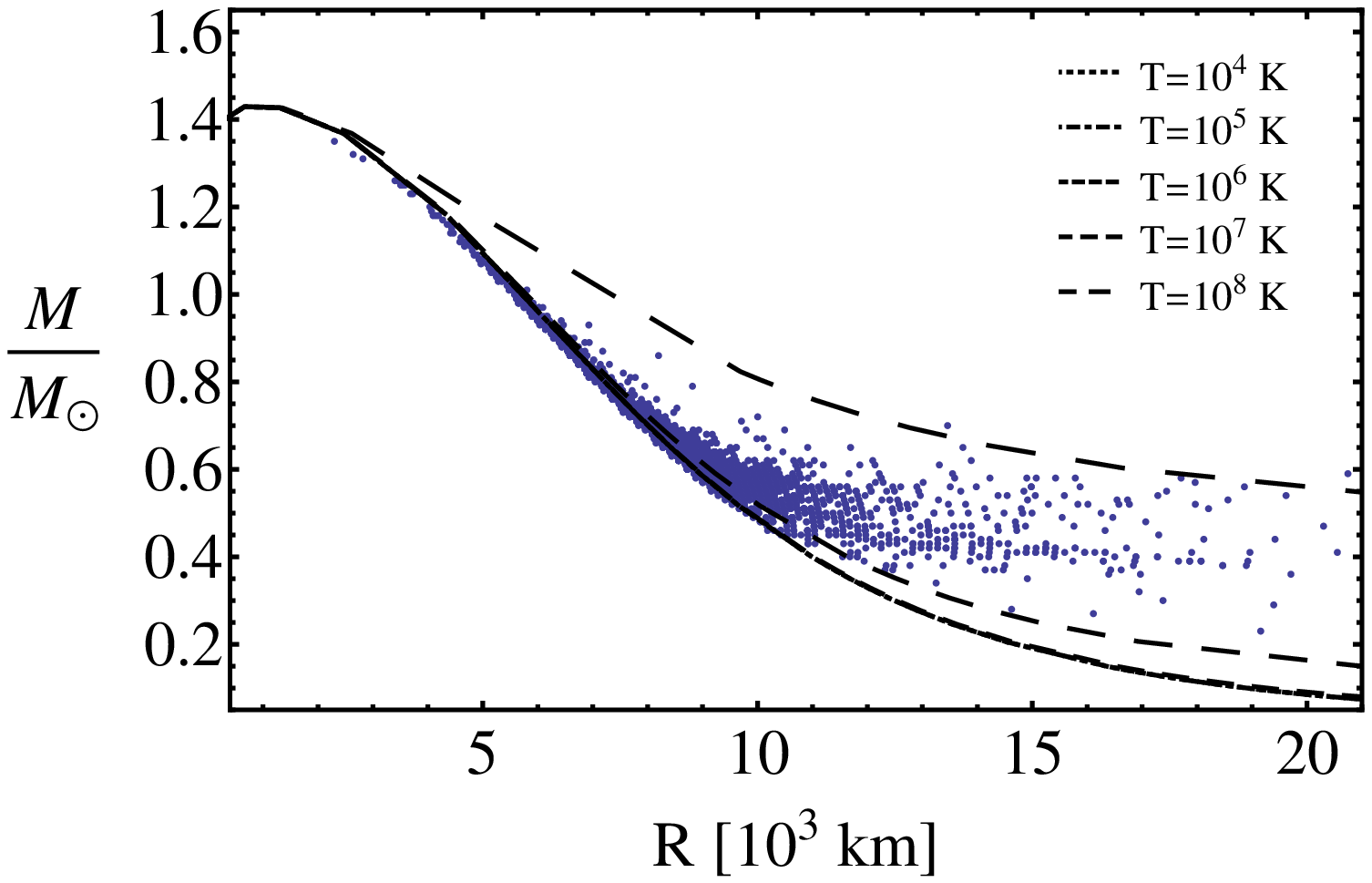}
\caption{Mass-radius relation of uniformly rotating WDs (the average radius is defined as $\left<R\right>=(1/3) \left(R_{p}+2R_{e}\right)$, where $P_{p}$ is the polar radius, $P_{e}$ is the equatorial radius) obtained with the Chandrasekhar EoS for T = 0 K case (left panel) and mass-radius relation of static WDs for selected finite temperatures from T =$10^4$ K to T = $10^8$ K  (right panel) and their superposition with the estimated masses and radii of WDs taken from the SDSS 4 (blue dots).}\label{fig:vs}
\end{center}
\end{figure}

In this work we construct the mass-radius relation for white dwarfs (WDs) using the Chandrasekhar equation of state (EoS) within general relativity (GR). First we perform computations for zero temperature uniformly rotating WDs at the mass shedding limit within the Hartle formalism\cite{bosh2013}. Afterwards we superpose our results with the estimated mass-radius relations obtained from the Sloan Digital Sky Survey Data Release 4 (SDSS 4) by Tremblay et al.\cite{tremb01} As one can see on the left panel of Fig.~\ref{fig:vs} the difference between theory and observations is quite noticeable.

In order to overcome this problem S. M.~de Carvalho et al.\cite{car01} proposed to include the finite-temperature effects in the EoS. Following this idea we performed similar analysis for static WDs at finite-temperatures by solving the Tolman-Oppenheimer-Volkoff equation (see Ref.~\refcite{bosh01} for details). The results of Ref.~\refcite{bosh01} are shown on the right panel of Fig.~\ref{fig:vs}. As one can see the only inclusion of the temperature effects on the EoS and on the structure of the WD leads to a mass-radius relation in better agreement with the observational data with respect to the only inclusion of the rotation effects.

It should be stressed that from the observations usually one infers the effective surface temperature and the surface gravity of WDs. All the rest parameters are estimated by using certain models. However there also exist techniques to measure the masses of WDs in close eclipsing binaries. The data inferred from close binaries are more reliable. Therefore in order to perform more realistic computations one needs to take into account the effects of rotation and temperature together for selected WDs with known parameters. Only after one can make further predictions. This issue is out of the scope of the present work and will be considered elsewhere.

In conclusion, we calculated the masses and radii of cold rotating and hot static WDs in GR i.e. the effects of the rotation and finite-temperature have been considered separately. We compared and contrasted our results with the estimated data from the observations of WDs. Our results cover all the data. For more detailed analysis one needs to consider both effects together and work with more or less model-independent data.

The work was partially supported by Grant No. 3101/GF4 IPC-11/2015 of the MES of the RK.

\end{document}